\documentclass[final,3p,times]{elsarticle}

\usepackage{geometry}
\usepackage{amssymb}

\usepackage{graphicx}
\usepackage{subfigure}
\usepackage{float}
\usepackage{amsmath}
\usepackage{color}
\usepackage{multirow}

\biboptions{sort&compress}

\begin{document}

\begin{frontmatter}



\title{A dynamic subgrid-scale modeling framework for large eddy simulation using approximate deconvolution}



\author{Romit Maulik}
\ead{romit.maulik@okstate.edu}
\author{Omer San\corref{cor1}}
\ead{osan@okstate.edu}
\cortext[cor1]{Corresponding author.}

\address{School of Mechanical and Aerospace Engineering, Oklahoma State University, Stillwater, Oklahoma 74078, USA}

\begin{abstract}

We put forth a dynamic modeling framework for sub-grid parametrization of large eddy simulation of turbulent flows based upon the use of the approximate deconvolution procedure to compute the Smagorinsky constant self-adaptively from the resolved flow quantities. Our numerical assessments for solving the Burgers turbulence problem shows that the proposed approach could be used as a viable tool to address the turbulence closure problem due to its flexibility.
\end{abstract}


\end{frontmatter}


\section{Introduction}
\label{}

Turbulent flows are encountered in a variety of engineering and geophysical systems involving a wide range of spatial and temporal scales. In a direct numerical simulation (DNS), the full spectra of turbulence should be resolved down to the Kolmogorov scale where the smallest feature of the motion is captured. The resolution requirements of small scale turbulence, however, are computationally prohibitive to fully resolve for all associated scales. Large eddy simulation (LES) aims to reduce this computational complexity and has been proven to be a promising approach for calculations of complex turbulent flows \citep{boris1992new,galperin1993large,lesieur1996new,piomelli1999large,meneveau2000scale}. Allowing much coarser spatial meshes, LES is designed to resolve the most energetic large scales of the turbulent motion while modeling small scales.

The LES equations are derived formally by applying a low-pass spatial filter to the governing equations of turbulent motion and can be considered a weak form or regularized form of the governing equations to remove the resolution requirement of small-scale turbulence. The nonlinear nature of the governing equations leads to the well-known closure problem in LES: sub-grid scale (SGS) parameterization where interactions between the small scales and the large ones need to be modeled. In the past few decades there has been a substantial effort on developing LES closure models using physical or mathematical arguments \citep{sagaut2006large,berselli2006mathematics}.

Being consistent with Kolmogorov's ideas on energy cascading, the \emph{eddy viscosity} approach is the foundation of the functional turbulent closure models. This approach yields one of the most celebrated closure models -- the Smagorinsky model \citep{smagorinsky1963general} -- which assumes that the eddy viscosity is computed from the resolved strain rate magnitude and a characteristic mixing length scale which is assumed to be proportional to the filter width via a modeling parameter usually called the Smagorinsky constant. Applications of the Smagorinsky model to various problems have revealed that the constant is not single-valued and varies depending on resolution and flow characteristics \citep{galperin1993large}. A major advance took place with the development of a dynamic model proposed by Germano et al. \cite{germano1991dynamic} and its modified form by Lilly \cite{lilly1992proposed} in which the Smagorinsky constant is self-adaptively determined from the simulation using a low-pass spatial test filter. This has made the eddy viscosity LES closure models more widely applicable in many fields (e.g., see \cite{piomelli1999large} and \cite{meneveau2000scale}). Layton \cite{layton2016energy} recently analyzed the dissipative behaviour of the Smagorinsky model and addressed the dynamic parameter selection as an important open problem, which is the topic of focus in the present study.

Alternatively, explicit filtering based structural closure models have been presented to treat the LES closure problem \citep{stolz1999approximate,lund2003use,bogey2006large,mathew2006new,bull2016explicit}. Among them Stolz and Adams \cite{stolz1999approximate} proposed {\em approximate deconvolution (AD)} framework using an iterative image deblurring technique \cite{biemond1990iterative} to estimate the unfiltered small scale contributions from the filtered flow variables by utilizing the Van Cittert iterations \citep{germano2009new,layton2012approximate,berselli2012convergence,germano2015similarity}. The AD method has been used successfully in the LES of turbulent flows (e.g., see \citep{san2016analysis} and references therein for a recent discussion).

In this note, we put forth a dynamic approach for SGS modeling of LES based upon using the AD process to estimate the eddy viscosity coefficient in a self-adaptive manner during simulations. The approach presented here is fundamentally different from the mixed methods where the AD model is coupled with the eddy viscosity models by adding a dynamically computed source term to increase the stability of the AD model \citep{gullbrand2003effect,habisreutinger2007coupled} and it is also different from the dynamic mixed scale-similarity models \citep{zang1993dynamic,sarghini1999scale,bouffanais2007large}. In our proposed framework, test filtering process of the standard dynamic model is replaced by the AD procedure and it is shown that the order of accuracy increases by increasing the number of the Van Cittert iterations. A posteriori error analysis is presented for solving the Burgers turbulence problem on various grid resolutions. Although the Burgers turbulence is a limiting case, it has been used to perform assessments for turbulence closure models \cite{love1980subgrid,adams2002subgrid,de2002sharp}. We assume that general conclusions will still hold since it retains some important properties of the Navier-Stokes equations \citep{falkovich2006lessons}.

\section{Dynamic modeling framework based upon approximate deconvolution}
\label{}
In this section, we present the proposed dynamic model for the Burgers equation, a simple prototype having the same quadratic nonlinearity. Its extension to the Navier-Stokes equation is straightforward. The viscous Burgers equation in the conservative form reads
\begin{align} \label{eq:ge}
 \frac{\partial u}{\partial t} + \frac{1}{2}\frac{\partial (u^2)}{\partial x} = \nu \frac{\partial^2 u}{\partial x^2},
\end{align}
where $u$ is the velocity, and $\nu$ is the kinematic viscosity. The filtered equations of motion for LES computations can be obtained by performing a low-pass filtering in the following form
\begin{align} \label{eq:fge}
 \frac{\partial \bar{u}}{\partial t} + \frac{1}{2}\frac{\partial (\bar{u}^2)}{\partial x} = \nu \frac{\partial^2 \bar{u}}{\partial x^2} + S,
\end{align}
where the overbar represents the filtered flow quantities, and $S$ is the SGS stress term to account for the effects of the truncated small scales to the resolved scales and can be formally written as
\begin{align} \label{eq:tau}
S =  \frac{1}{2}\frac{\partial (\bar{u}^2)}{\partial x}-\frac{1}{2}\overline{\frac{\partial(u^2)}{\partial x}},
\end{align}
in which an LES closure model is required to approximate unfiltered quantities in the second term of right-hand-side of Eq.~(\ref{eq:tau}) from the resolved flow variables. Assuming the dissipation of kinetic energy at sub-grid scales to be analogous to molecular diffusion, Smagorinsky's eddy viscosity model approximates the SGS term as
\begin{align} \label{eq:smag}
S = \frac{\partial}{\partial x}\left(\nu_e \frac{\partial \bar{u}}{\partial x} \right),
\end{align}
where $\nu_e$ is the eddy viscosity which is given by
\begin{align} \label{eq:ev}
\nu_e = (C_S \delta)^2 \left|\frac{\partial \bar{u}}{\partial x}\right|,
\end{align}
in which $C_S$ is the Smagorinsky constant and $\delta$ is formally defined by the filter width and usually set to the representative mesh size. Following the classical dynamic model procedure \citep{germano1991dynamic,lilly1992proposed}, the modeling constant $(C_S \delta)^2$ can be computed self-adaptively by defining a test filter (e.g., see \citep{fauconnier2009family} for the derivation using the Burgers equation). Here, we purpose a different approach to determine the modeling constant based upon using the AD procedure. If $\vartheta$ is assumed to be an approximately recovered value for the unfiltered quantity $u$, the Van Cittert iterations can be used to estimate $\vartheta$ from $\bar{u}$:
\begin{eqnarray}\label{eq:adm}
    \vartheta^0 &=& \bar{u} \nonumber\\
    \vartheta^i &=& \vartheta^{i-1} + \beta(\bar{u} - G \ast \vartheta^{i-1}), \quad i = 1,2,3,...,Q,
\end{eqnarray}
where $G$ is the low-pass spatial filtering operator (i.e., $G \ast u = \bar{u}$) and this approach yields increasingly accurate reconstructions with increasing $Q$, the order of the Van Cittert iteration \citep{dunca2006stolz}. Here, $\beta$ is the over-relaxation parameter and it can be shown that $0<\beta\leq2$ when the transfer function of the spatial filter $0\leq T(k)\leq1$ (e.g, the reader is referred to \cite{biemond1990iterative} for convergence characteristics of the Van Cittert iteration). We have set $\beta=2$ in our computations. Then, Eq.~(\ref{eq:tau}) can be approximated as
\begin{align} \label{eq:ad}
S =  \frac{1}{2}\frac{\partial (\bar{u}^2)}{\partial x}-\frac{1}{2}\overline{\frac{\partial(\vartheta^2)}{\partial x}},
\end{align}
where $\vartheta \approx u$. Using Eq.~(\ref{eq:ev}) in Eq.~(\ref{eq:smag}), the following relationship should hold true
\begin{align} \label{eq:equ}
\frac{1}{2}\frac{\partial (\bar{u}^2)}{\partial x}-\frac{1}{2}\overline{\frac{\partial(\vartheta^2)}{\partial x}} = \frac{\partial}{\partial x}\left((C_S \delta)^2 \left|\frac{\partial \bar{u}}{\partial x}\right|\frac{\partial \bar{u}}{\partial x} \right),
\end{align}
when we use the AD procedure given by Eq.~(\ref{eq:adm}) in Eq.~(\ref{eq:ad}). Factoring the model coefficient, Eq.~(\ref{eq:equ}) can be recast in the following form
\begin{align} \label{eq:min}
     L = (C_S \delta)^2 M,
\end{align}
where
\begin{align} \label{eq:lm}
     L = \frac{1}{2}\frac{\partial (\bar{u}^2)}{\partial x}-\frac{1}{2}\overline{\frac{\partial(\vartheta^2)}{\partial x}}, \quad \quad  M = \frac{\partial}{\partial x}\left( \left|\frac{\partial \bar{u}}{\partial x}\right|\frac{\partial \bar{u}}{\partial x} \right).
\end{align}

Following \citep{lilly1992proposed}, the coefficient $(C_S \delta)^2$ in Eq.~(\ref{eq:min}) can be obtained by minimizing the average square error $\langle E^2 \rangle$, where the error is given by $E=L-(C_S \delta)^2 M$. Differentiating the mean square error with respect to the model parameter $(C_S \delta)^2$, we can minimize the square of the error when
\begin{align}
    (C_S \delta)^2 = \frac{\langle L M \rangle}{\langle M^2 \rangle},
\end{align}
where the averaging operator is expressed by $\langle f \rangle = \frac{1}{2\pi} \int_{0}^{2\pi} f dx$ for any quantity $f$ in the present study. The definition of the spatial filter $G$ completes the formulation of the proposed dynamic model. Here, we use the following discrete filter \cite{san2016analysis}
\begin{equation}
\bar{f}_{j}= \frac{1}{64}(f_{j-3}-6f_{j-2}+15f_{j-1}+44f_{j}+15f_{j+1}-6f_{j+2}+f_{j+3}) ,
\label{eq:fil}
\end{equation}
where the complete attenuation happens for the largest wavenumber.

\begin{figure}[!t]
\centering
\mbox{
\subfigure{\includegraphics[width=0.47\textwidth]{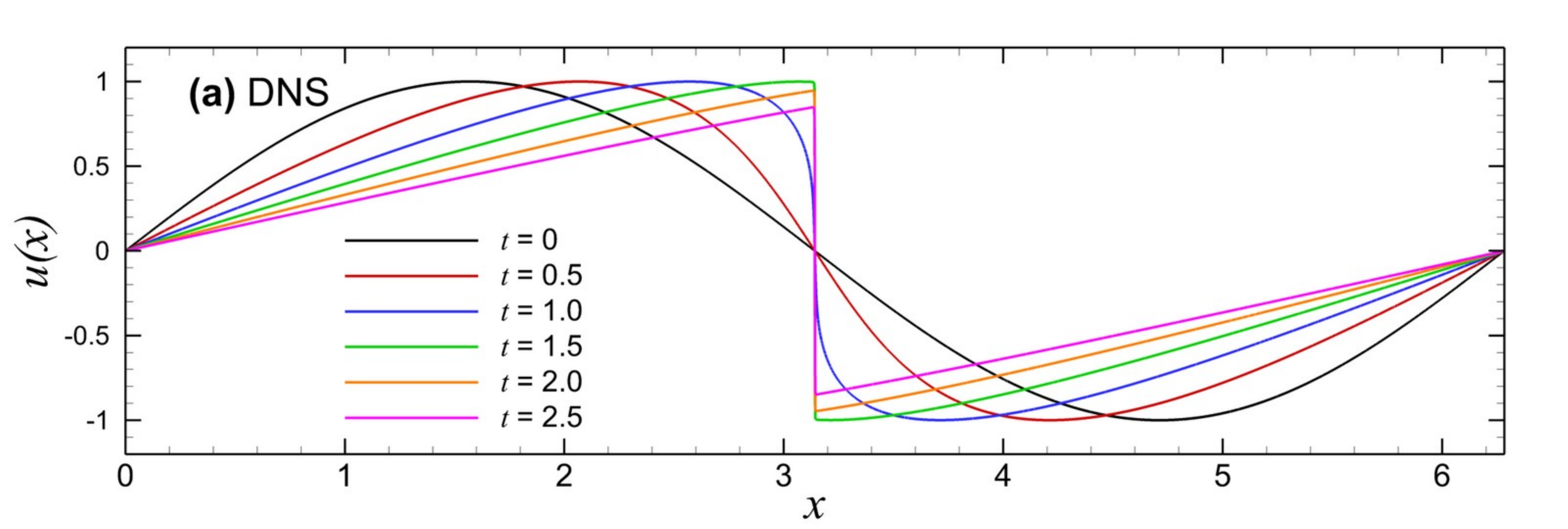}}
\subfigure{\includegraphics[width=0.47\textwidth]{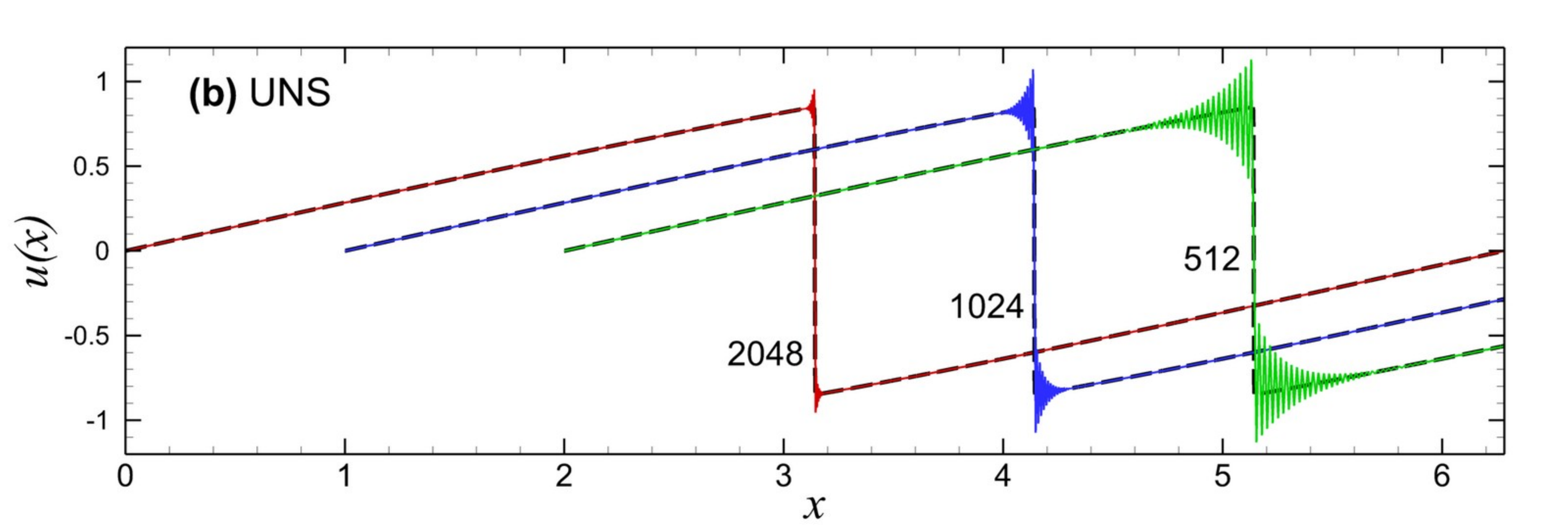}}
}\\
\mbox{
\subfigure{\includegraphics[width=0.47\textwidth]{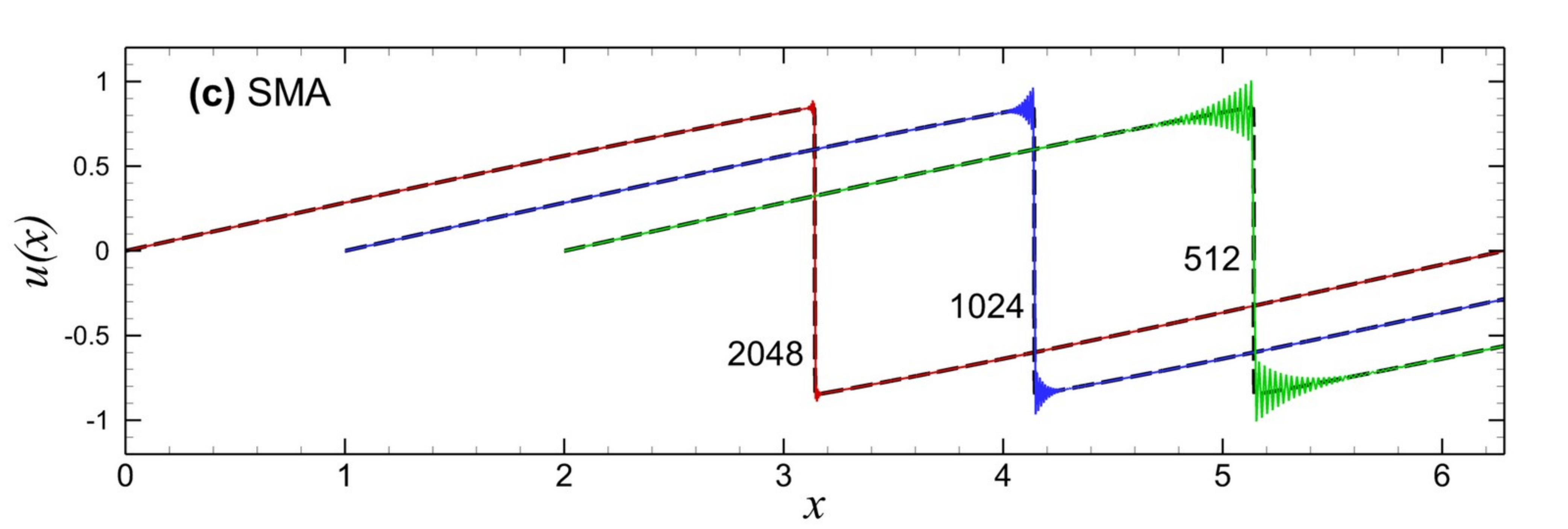}}
\subfigure{\includegraphics[width=0.47\textwidth]{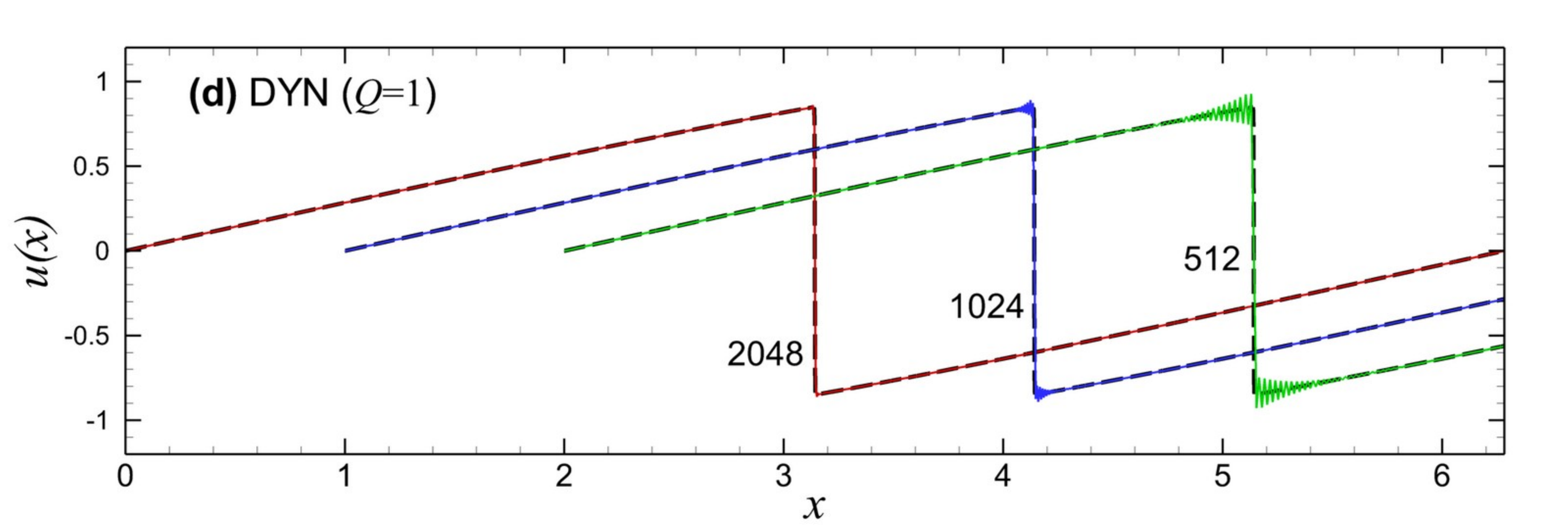}}
}\\
\mbox{
\subfigure{\includegraphics[width=0.47\textwidth]{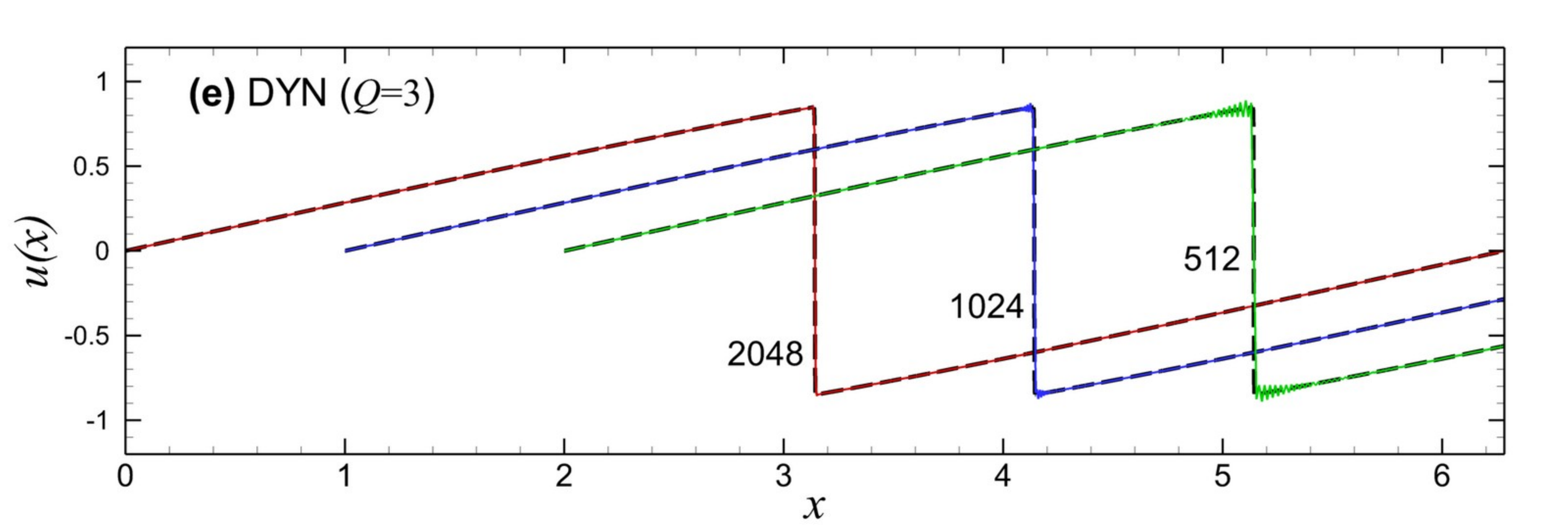}}
\subfigure{\includegraphics[width=0.47\textwidth]{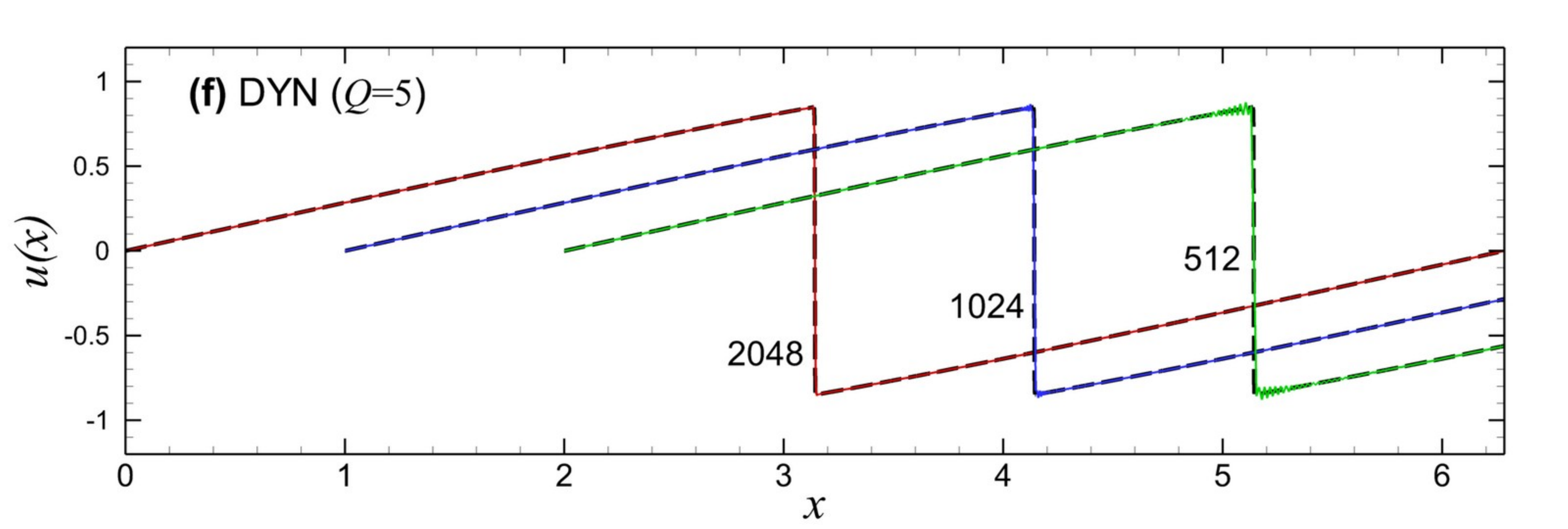}}
}\\
\caption{Simulation results for Burgers equation initiated by the single-mode sine wave: (a) time evolution by DNS with resolution of $N=32768$; (b) under-resolved numerical simulation (UNS), (c) Smagorisnky model (SMA) with $C_S=0.24$, and (d-f) the proposed dynamic model (DYN) with various order $Q$, using resolutions of $512$, $1024$, and $2048$ at time $t=2.5$ where the results for $512$ and $1024$ resolutions are shifted in $x$-axis for the purpose of illustration. The dashed lines in (b-f) represent the DNS data.}
\label{fig:sine}
\end{figure}

\section{Decaying Burgers turbulence simulations}
\label{}
In this section, we present our results for two test cases: (i) the shock formation problem initiated by a single-mode sin wave, $u(x,0)=sin(x)$, and (ii) the decaying Burgers turbulence problem initiated by a specified energy spectrum considering the 64 sample initial conditions associated with randomly generated phases. Both problems are considered in a domain of $x \in [0,2\pi]$ with periodic boundary conditions using $\nu = 5 \times 10^{-4}$. Time integration for our discrete system is carried out using a third-order accurate total variation diminishing Runge-Kutta scheme with $\Delta t= 1\times10^{-5}$, and the sixth-order central compact difference schemes were used for the spatial discretizations to ensure a minimal numerical error. Further details on the initial energy spectrum and numerical methods can be found in \cite{san2016analysis}.

\begin{figure}[!t]
\centering
\includegraphics[width=0.45\textwidth]{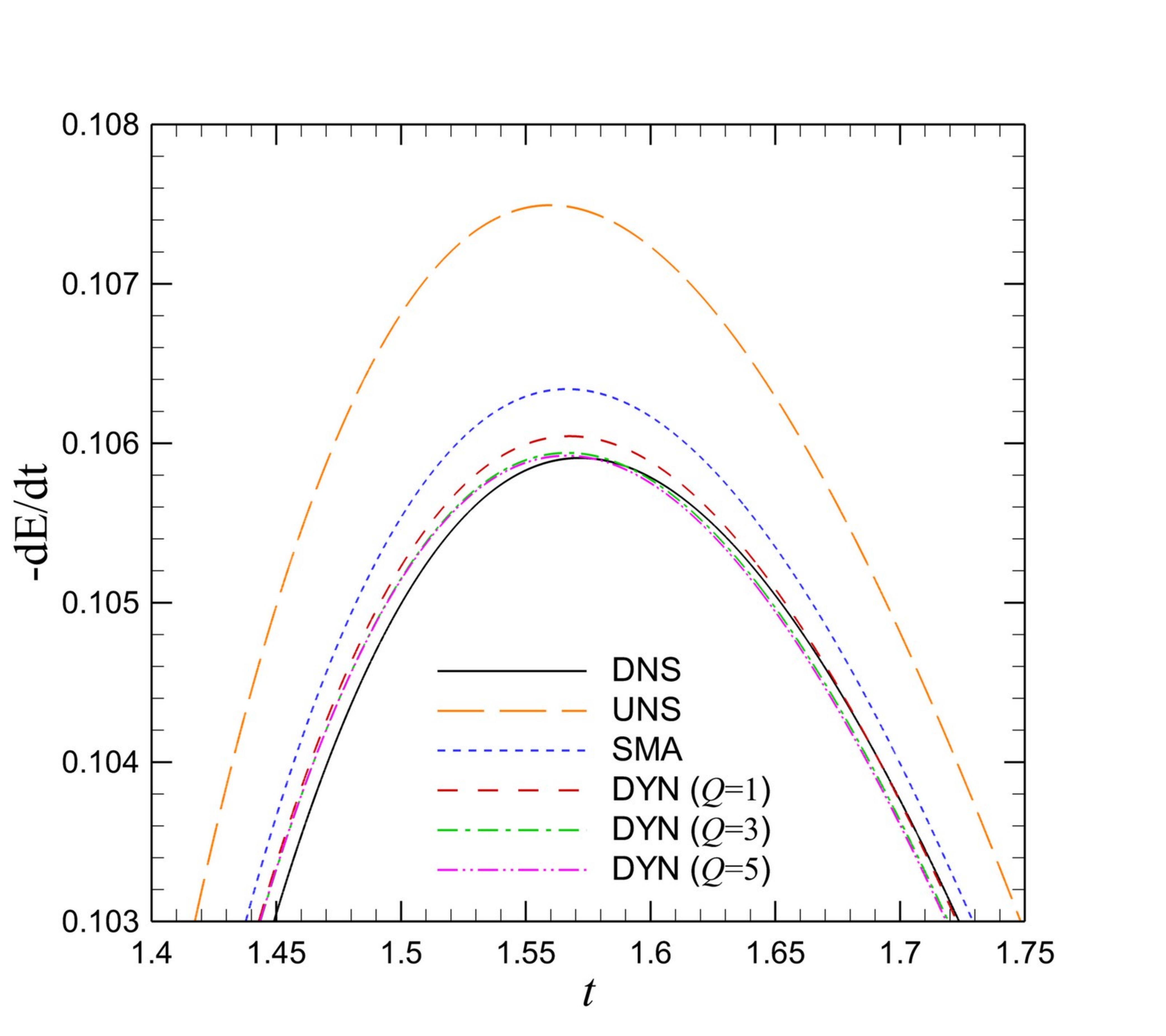}
\caption{A close-up comparison for the time evolution of the dissipation rate for the shock formation problem initiated by a single-mode sine wave.}
\label{fig:srate}
\end{figure}

Fig.~\ref{fig:sine} shows simulation results for the first problem initiated by the single-mode sin wave. The time evolution of the wave propagation obtained by a DNS computation on the resolution of $N=32768$ is illustrated in Fig.~\ref{fig:sine}(a) demonstrating the shock formation at a time approximately equal to 1.55. After the shock developed, results at the time $t=2.5$ obtained by the under-resolved numerical simulation (UNS) where no-turbulence model is used, and the Smagorinsky model are presented in Fig.~\ref{fig:sine}(b) and Fig.~\ref{fig:sine}(c), respectively, with resolutions of $N=512$, $N=1024$, and $N=2048$, showing grid-to-grid oscillations, especially for the coarsest grid. On the other hand, as shown in Fig.~\ref{fig:sine}(d-f), the proposed dynamic model yields substantially better results with damping the grid-to-grid oscillations by increasing $Q$, the order of Van Cittert iteration, which also can be seen from Fig.~\ref{fig:srate} demonstrating the total dissipation rate $-dE/dt$ where $E=\int E(k)dk$ is the total kinetic energy.

\begin{figure}[!h]
\centering
\mbox{
\subfigure{\includegraphics[width=0.45\textwidth]{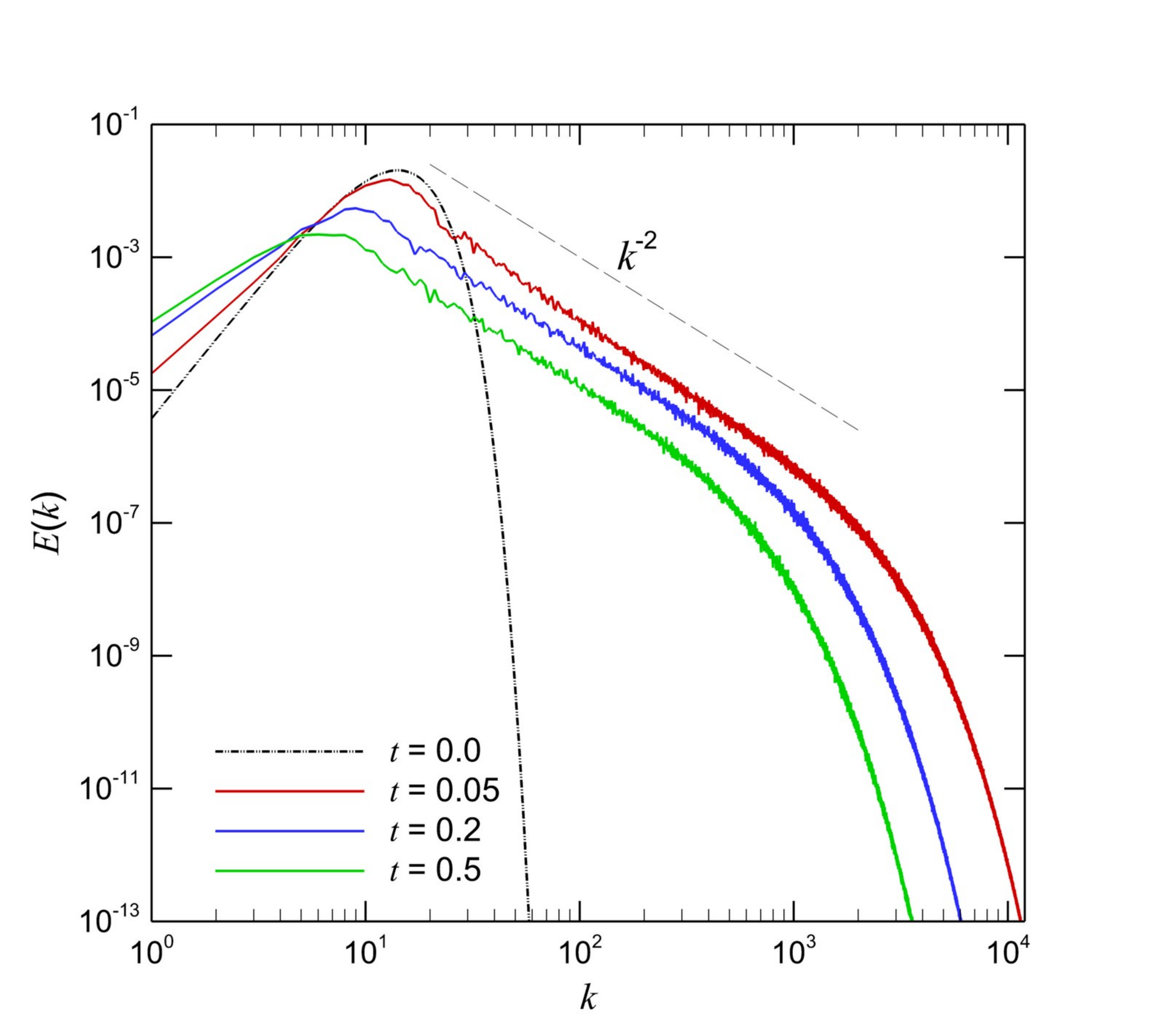}}
\subfigure{\includegraphics[width=0.45\textwidth]{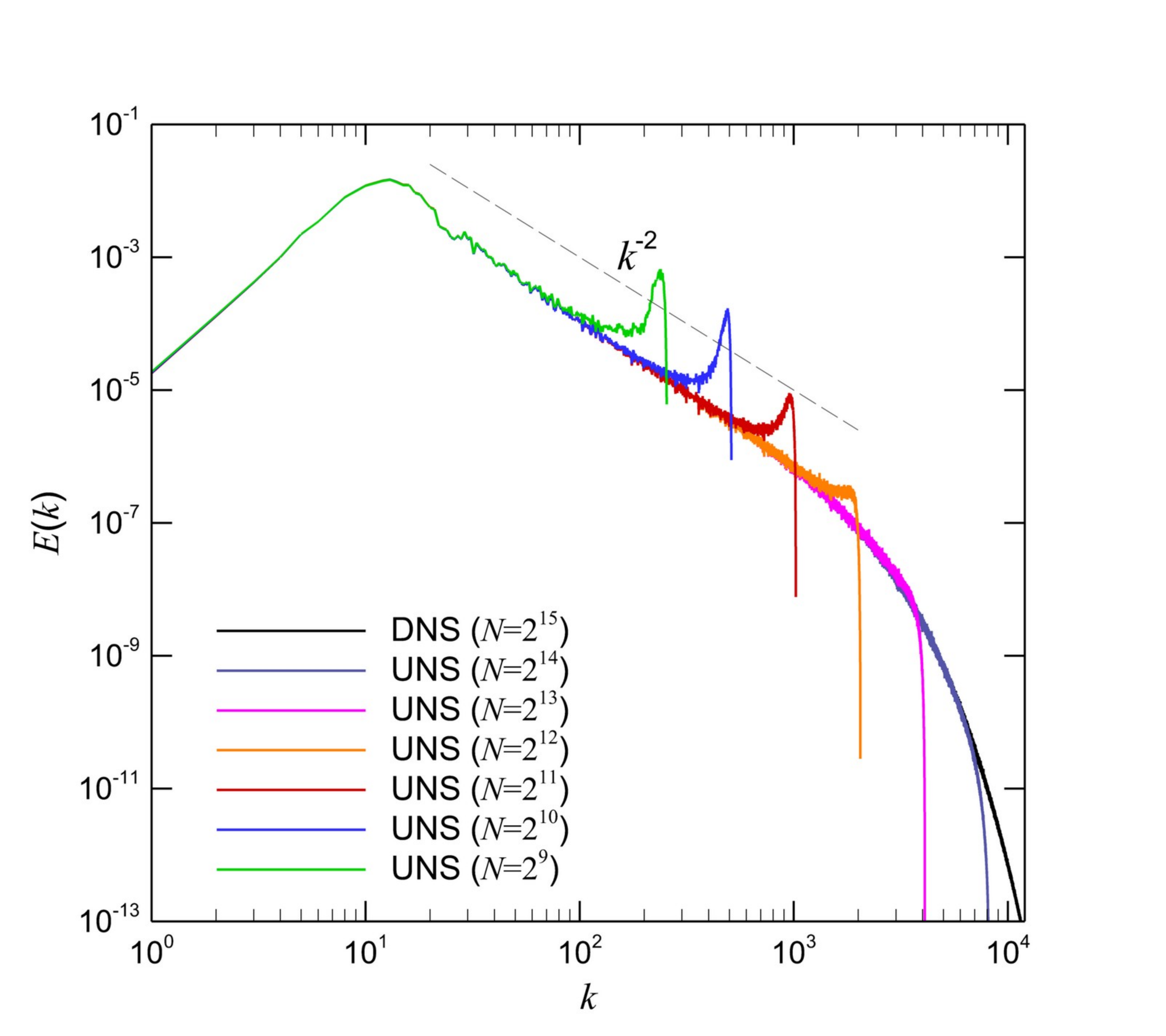}}
}\\
\caption{Time evolution of the Burgers turbulence problem based upon ensemble average of 64 sample fields obtained by DNS ($N=32768$) (left); and energy spectra at time $t=0.05$, obtained by various spatial resolutions ranging from $N=512$ to $N=32768$ (right).}
\label{fig:dns}
\end{figure}

\begin{figure}[!h]
\centering
\mbox{
\subfigure{\includegraphics[width=0.45\textwidth]{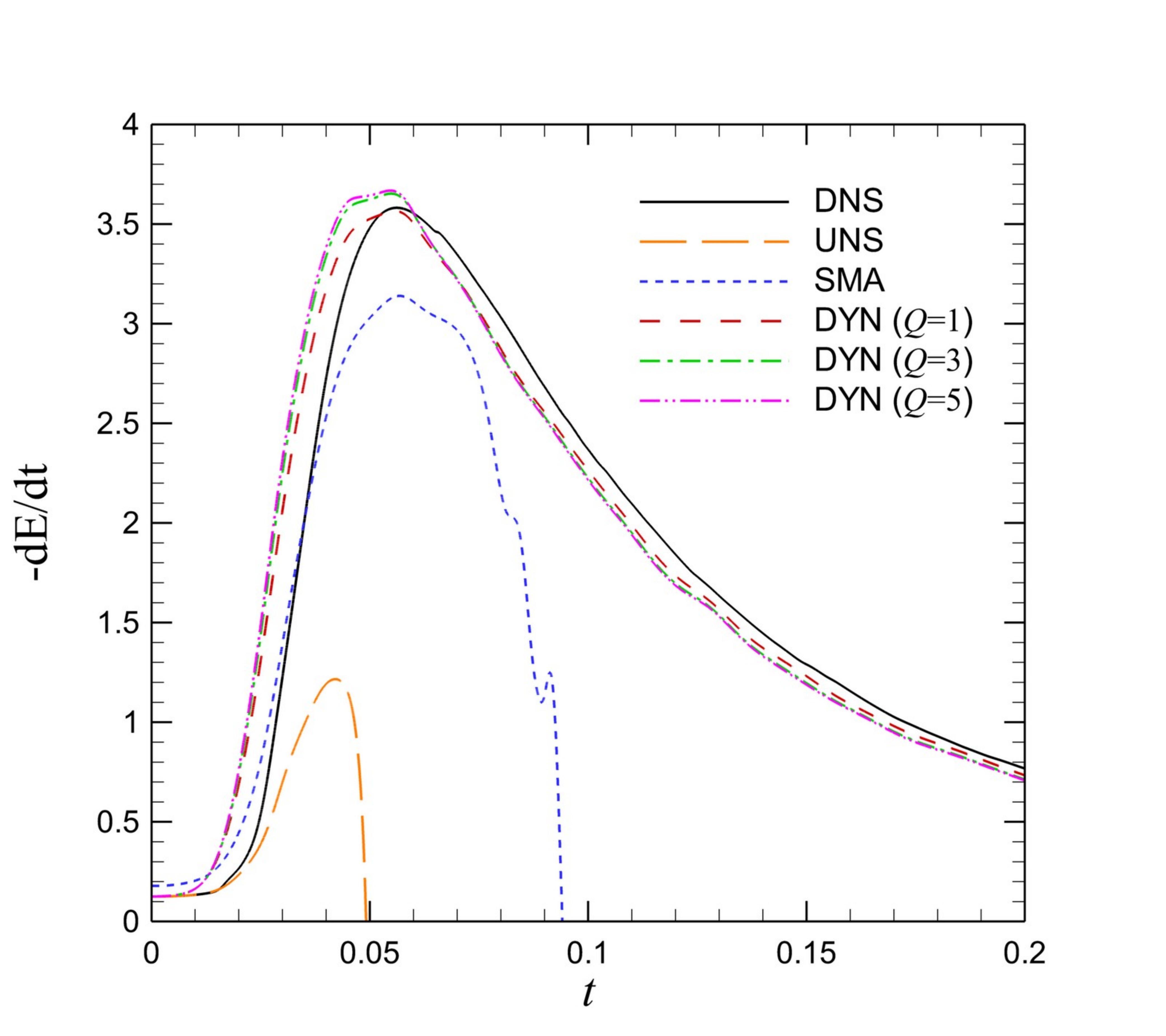}}
\subfigure{\includegraphics[width=0.45\textwidth]{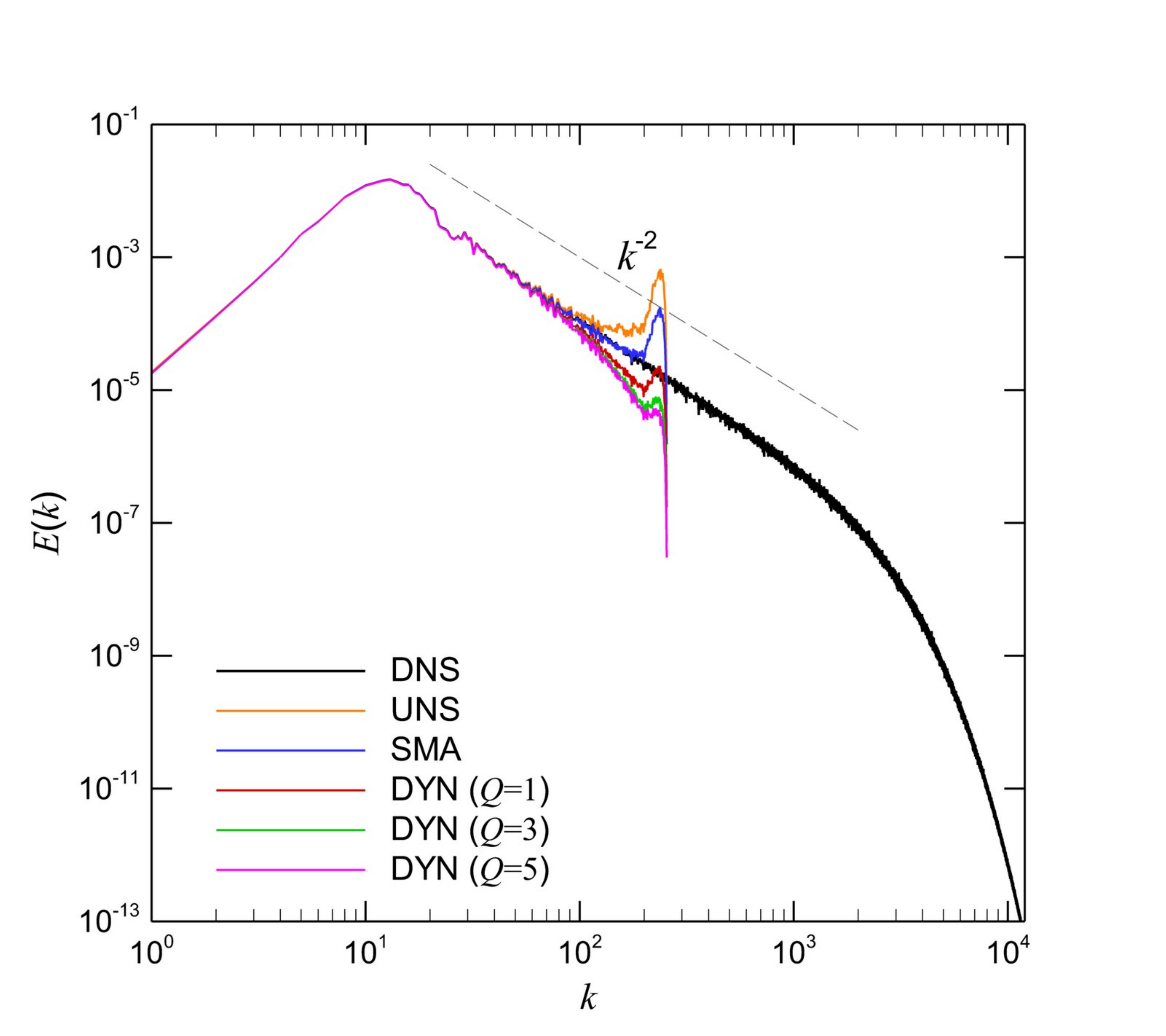}}
}\\
\caption{SGS modeling results obtained by using $N=512$ resolution: time evolution of the total dissipation rate (left), and energy spectra at time $t=0.05$ (right). Results from DNS and UNS computations are also included for comparison purposes.}
\label{fig:turb}
\end{figure}

Next, we present our results for the Burgers turbulence problem. Fig.~\ref{fig:dns}(a) illustrates the time evolution of the decaying Burgers turbulence based upon ensemble average of 64 sample fields, obtained by DNS with resolution of $N=32768$. Fig.~\ref{fig:dns}(b) shows the energy spectra at $t=0.05$ for DNS and UNS computations for different resolutions demonstrating the ideal $k^{-2}$ scaling at inertial range. It is clear from the figure that the resolution of $N=32768$ yields DNS data resolving all associated shock scales at $\nu = 5 \times 10^{-4}$. One must note that $t=0.05$ corresponds approximately to the time for the maximum dissipation rate of the problem. The energy pile-up close to grid cut-off scale is also evident at lower resolutions. In Fig.~\ref{fig:turb}, we present our LES analysis by using the coarsest resolution of $N=512$ illustrating the time evolution of the total dissipation rate and energy spectra at time $t=0.05$. Results obtained by DNS, UNS and the Smagorinsky model with $C_S=0.24$ are also included for comparison purposes. It is shown that the proposed dynamic model yields a dissipation rate with considerably higher accuracy and the energy pile-up can be effectively eliminated by increasing $Q$. Being consistent with the previous test case, increasing $Q$ further yields more dissipative results. This implies that more dissipative behaviour can be obtained by using a higher order Van Cittert iterative procedure, which would be crucial for solving LES equations on very coarse grid resolutions. We also found that results are negligibly improved by increasing the order of Van Cittert iterations beyond $Q=5$.

\section{Conclusion}
\label{}
We propose a dynamic framework for SGS modeling of LES of turbulent flows based upon the use of AD process to compute the Smagorinsky constant self-adaptively from the resolved flow quantities. Our proposed model replaces the test filtering procedure of the standard dynamic model with the AD procedure and can be considered more general by virtue of its flexibility. Although the presentation of the model is given by using the Burgers equation, its generalization to the Navier-Stokes equation is straightforward. Our first step in the numerical assessment for solving the Burgers turbulence problem shows that the proposed approach could represent a viable tool for sub-grid scale parametrization of turbulent flows by combining the AD process with the celebrated dynamic model.









\bibliographystyle{elsarticle-num}
\bibliography{references}

\end{document}